\documentclass[12 pt]{article}
\usepackage[left=1in, right=1in, top=0.75in, bottom=0.75in]{geometry}
\usepackage[utf8]{inputenc}
\usepackage{textcomp}
\usepackage{setspace}
\usepackage{graphicx}
\usepackage{lmodern} 
\usepackage{upgreek} 
\usepackage{chemformula}
\usepackage{authblk}
\usepackage{booktabs}
\usepackage{gensymb}
\usepackage{amsmath}
\usepackage{parskip}
\usepackage{caption}
\usepackage{titlesec}
\usepackage{wrapfig}
\usepackage{amssymb}
\usepackage{xcolor}

\usepackage[utf8]{inputenc}
\usepackage[T1]{fontenc}
\usepackage{siunitx}
\usepackage{booktabs}
\usepackage[version=4]{mhchem}
\DeclareSIUnit\angstrom{\text {Å}}

\titlespacing{\section}{0pt}{12pt plus 2pt minus 2pt}{12pt plus 2pt minus 2pt}
\usepackage{sectsty}
\sectionfont{\fontsize{14}{16}\selectfont}
\subsectionfont{\fontsize{12}{13}\selectfont}

\usepackage[backend=biber, style=chem-acs,articletitle=true,sorting=none]{biblatex}
\newcommand{\WSe}{1\ch{\emph{H}-WSe2}}
\newcommand{\TTaS}{1\ch{\emph{T}-TaS2}}

\title{\Large\textbf{Local interface effects modulate global charge order and optical properties of \TTaS/\WSe\ heterostructures}}

\author[1,$\dagger$]{Samra Husremovi\'{c}}
\author[1,$\dagger$]{Valerie S. McGraw}
\author[2]{Medha Dandu}
\author[1]{Lilia S. Xie}
\author[3]{Sae Hee Ryu}
\author[1]{Oscar Gonzalez}
\author[1]{Shannon S.\ Fender}
\author[1]{Madeline Van Winkle}
\author[2]{Karen C. Bustillo}
\author[4]{Takashi Taniguchi}
\author[5]{Kenji Watanabe}
\author[3]{Chris Jozwiak}
\author[3]{Aaron Bostwick}
\author[3]{Eli Rotenberg}
\author[2,7]{Archana Raja}
\author[6]{Katherine Inzani}
\author[1,7,8,*]{D. Kwabena Bediako}

\affil[1]{Department of Chemistry, University of California, Berkeley, CA 94720, USA}
\affil[2]{Molecular Foundry, Lawrence Berkeley National Laboratory, Berkeley, CA, USA}
\affil[3]{Advanced Light Source, Lawrence Berkeley National Laboratory, Berkeley, CA, 94720, USA}
\affil[4]{\textit{Research Center for Materials Nanoarchitectonics, National Institute for Materials Science,  1-1 Namiki, Tsukuba 305-0044, Japan}}
\affil[5]{\textit{Research Center for Electronic and Optical Materials, National Institute for Materials Science, 1-1 Namiki, Tsukuba 305-0044, Japan}}
\affil[6]{\textit{School of Chemistry, University of Nottingham, United Kingdom}}
\affil[7]{\textit{Kavli Energy NanoScience Institute, Berkeley, CA 94720, USA}}
\affil[8]{\textit{Chemical Sciences Division, Lawrence Berkeley National Laboratory, Berkeley, CA 94720, USA}}
\affil[*]{Correspondence to: bediako@berkeley.edu}
\affil[$\dagger$]{These authors contributed equally to this work}

\date{}
\begin{document}
\maketitle

\doublespacing

\textit{Abstract}
\break
\small{\TTaS\ is a layered charge density wave (CDW) crystal exhibiting sharp phase transitions and associated resistance changes. These resistance steps could be exploited for information storage, underscoring the importance of controlling and tuning CDW states. Given the importance of out-of-plane interactions in \TTaS, modulating interlayer interactions by heterostructuring is a promising method for tailoring CDW phase transitions. In this work, we investigate the optical and electronic properties of heterostructures comprising \TTaS\ and monolayer \WSe. By systematically varying the thickness of \TTaS\  and its azimuthal alignment with \WSe, we find that intrinsic moir\'e strain and interfacial charge transfer introduce CDW disorder in \TTaS\ and modify the CDW ordering temperature. Furthermore, our studies reveal that the interlayer alignment impacts the exciton dynamics in \WSe, indicating that heterostructuring can concurrently tailor the electronic phases in \TTaS\ and the optical properties of \WSe. This work presents a promising approach for engineering optoelectronic behavior of heterostructures that integrate CDW materials and semiconductors.}

\newpage

\section*{Introduction}
Charge density waves (CDWs) are correlated electronic systems that exhibit periodic lattice distortions, which template modulations in charge density \cite{gruner_density_1994}. This intrinsic charge ordering strongly influences the measured resistance, rendering CDW phase transitions promising for information storage and processing applications \cite{xu2021topical,mihailovic2021ultrafast}. \TTaS, a layered van der Waals (vdW) crystal, has garnered attention as a platform for next-generation electronics due to its high-temperature CDW phase transitions accompanied by sharp resistance changes. Below $\sim$ 500 K \cite{geremew2019bias,sung2024endotaxial,fazekas1979electrical}, periodic star-shaped lattice distortions emerge in \TTaS\ layers. As the in-plane tiling of clusters evolves with temperature, a series of metal-to-insulator transitions occur \cite{Hovden2016,husremovic2023encoding,wilson1975charge,scruby1975role}. The metallic incommensurate CDW (IC-CDW) phase, observed between 500 K and 350 K, exhibits sparse and irregular cluster arrangements. In contrast, the nearly commensurate phase (NC-CDW), typically seen between 350 K and 180 K, is marked by the formation of locally insulating domains of cluster superlattices separated by metallic discommensuration networks \cite{Hovden2016,scruby1975role,spijkerman1997x,yamamoto1983hexagonal,yamada1977origin,wilson1990solution}. Below $\sim$ 180 K, the discommensuration networks disappear, resulting in the insulating commensurate phase (C-CDW) with a $\sqrt{13}a \times \sqrt{13} a$ periodicity, where $a$ is the in-plane lattice constant of \TTaS\ \cite{Hovden2016,husremovic2023encoding}.

Although \TTaS\ is typified by in-plane lattice distortions, the out-of-plane stacking of CDW clusters also changes across the CDW phases and strongly influences the electronic properties \cite{robbins1980x, ritschel2015orbital, stahl2020collapse,scruby1975role,tanda1984x,butler2020mottness,ritschel2018stacking}. Consequently, modulating interlayer interactions of \TTaS\ through heterostructure assembly is a compelling means of fine-tuning its properties \cite{martino_preferential_2020}. Recent studies have demonstrated the potential of this approach by stabilizing a robust, room-temperature C-CDW phase in polytype heterostructures of \TTaS\ and \ch{\emph{H}-TaS2} grown using moderate thermal annealing of \TTaS\ \cite{sung2022two,sung2024endotaxial,husremovic2023encoding}. However, the use of vdW assembly to integrate \TTaS\ and other layered CDW crystals into heterostructures remains largely unexplored \cite{chen2018interlayer,wang2019modulating,zhao_tuning_2017} despite its theoretical promise \cite{Li_2019, Goodwin_2022} and the structural tunability that this approach offers. Although experimental studies are limited, existing work suggests that heterostructuring CDW materials can yield emergent phenomena, such as CDW-activated excitons in TiSe$_2$/MoSe$_2$ heterostructures\cite{Joshi_2022}.

In vdW assembly, dissimilar layered crystals can be stacked without lattice matching restrictions, enabling fine control over interfacial hybridization and charge transfer characteristics---both of which play a defining role in CDW phases \cite{Wei_manipulating2017,dreher2021proximity}. Compared to other methods for controlling these properties, such as atomic/ionic doping \cite{liu_superconductivity_2013,jung_control_2023,li2012fe,chen_influence_2015}, vdW assembly introduces fewer atomic defects and inhomogeneities into the system \cite{nessralla2023modulating,bediako2018heterointerface}. The characteristics of vdW heterostructures can be further modulated by the presence of interfacial moir\'e patterns, formed when vertically stacked vdW crystals possess an azimuthal misorientation (twist) and/or differences in lattice constants \cite{Cao2018, seyler2019signatures, mak2022semiconductor, he2021moire}. At small twist angles and lattice mismatch values, the lattice reconstructs to minimize the area of high-energy stacking domains, which reduces the interlayer stacking energy but introduces strain \cite{kazmierczak_strain_2021, VanWinkle2023}. This intrinsic moir\'e strain has been shown to template the CDW domains in 1\ch{\emph{T}-TiTe2}/1\ch{\emph{T}-TiSe2} \cite{zhao_moire_2021}. Engineering intrinsic moir\'e strain could be a powerful tuning knob for \TTaS, due to its reported responsiveness to extrinsic strain \cite{zhao_tuning_2017,nicholson_gap_2024,gan2016strain}.  Nevertheless, studies on \TTaS\ heterostructures remains relatively limited \cite{chen2018interlayer,wang2019modulating}, with the absence of twist-dependent measurements hindering the understanding of how charge transfer, strain, and hybridization contribute to ensemble CDW properties.

In this work, we investigate the optical and electronic properties of heterostructures comprising vertically stacked nanothick crystals of \TTaS\ and monolayer \WSe. By systematically varying the thickness of \TTaS\ and its azimuthal misalignment with \WSe, we observe that intrinsic moir\'e relaxation strain and interfacial charge transfer dictate the global CDW ordering in few-layer \TTaS. Specifically, local charge transfer, which is independent of the twist angle, induces increased CDW disorder and lowers the IC-CDW--NC-CDW transition temperature. This trend notwithstanding, we consistently measure a more pronounced CDW suppression for heterostructures with small interlayer twist angles. We attribute this effect to the intralayer strain from moir\'e reconstruction, evident in our transmission electron microscopy (TEM) studies and confocal Raman spectroscopy of \WSe. Our study additionally extends to the optoelectronic characteristics of the \WSe\ layers, where we find that the exciton binding energy and exciton dissociation dynamics are affected by the interlayer twist angle between \WSe\ and \TTaS. These results demonstrate a means to concurrently engineer CDW phases of \TTaS\ and the optical properties of \WSe\ through targeted adjustments of local interfacial interactions. More generally, this work illustrates a promising route toward deterministic engineering of electronic and optical characteristics of heterostructures integrating atomically thin CDW materials and semiconductors.

\section{Results and Discussion}
\subsection{Electronic properties of \TTaS/\WSe\ heterostructures}

\begin{figure}[!htbp]
    \centering
    \includegraphics[width=\textwidth]{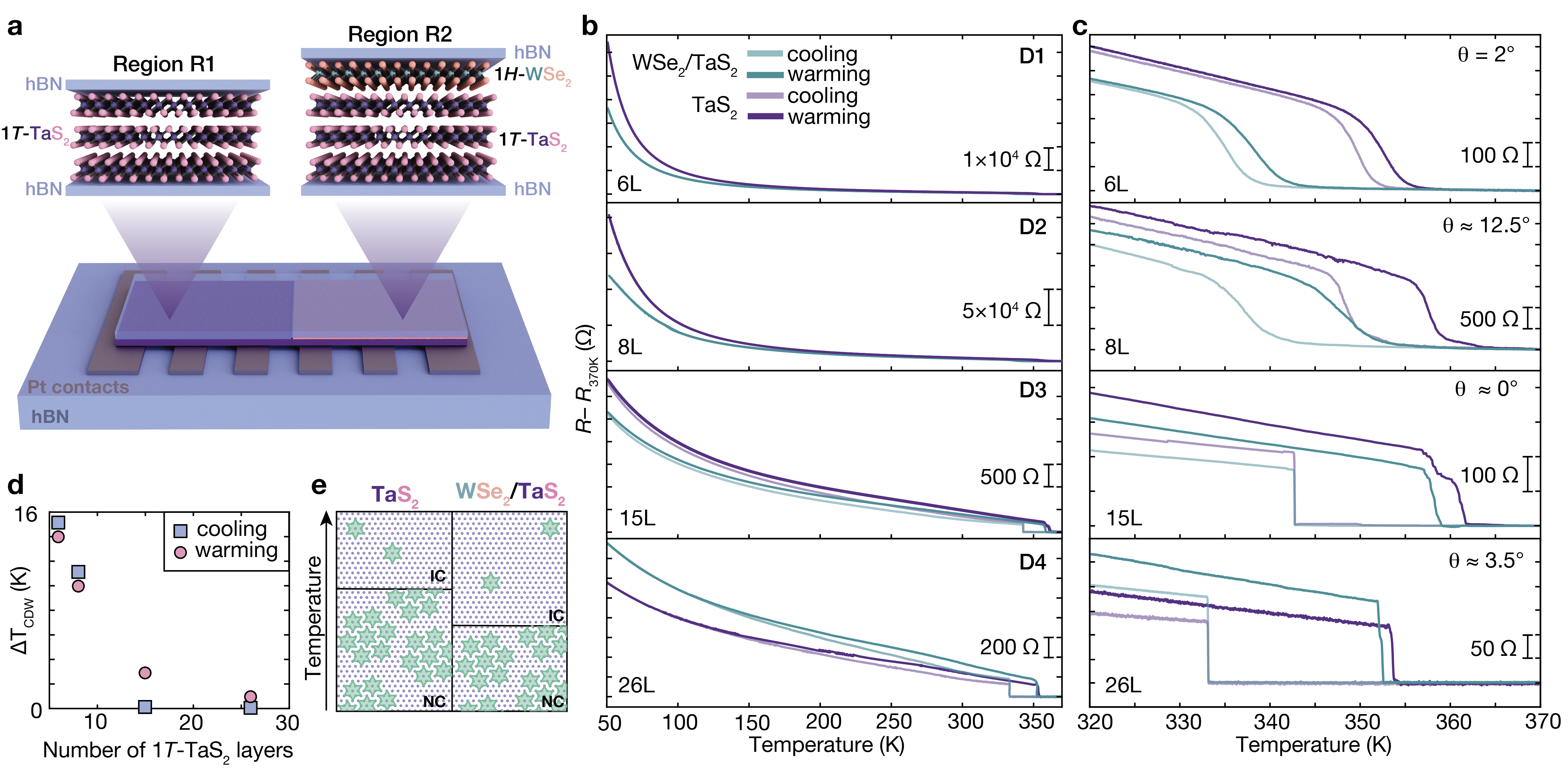}
    \caption[Electronic measurements of CDW phase transitions in \WSe/\TTaS\ heterostructures.] {Electronic measurements of CDW phase transitions in \WSe/\TTaS\ heterostructures. \textbf{(a)} Schematic of \WSe/\TTaS\ heterostructure device for transport measurements. Each \TTaS\ crystal is fully encapsulated in hBN and partially covered with monolayer \WSe, yielding a \TTaS--only region (region R1) and a \TTaS/\WSe\ region (region R2). Electrical contact with the \TTaS\ crystal is made using pre-patterned Pt contacts fabricated on hBN. \textbf{(b)} Temperature-dependent longitudinal resistance ($R_{xx}$) of devices fabricated according to (a) containing 6-, 8-, 15-, and 26-layer \TTaS. These devices are labeled D1--D4, respectively. The twist angle ($\theta$) between \TTaS\ and \WSe\ is marked for each device. For D1, $\theta$ is derived from TEM diffraction, while for D2--D4, $\theta$ is inferred from the edge alignment of \TTaS\ and \WSe\ in optical micrographs. \textbf{(c)} High-temperature (320--370 K) section of (b). \textbf{(d)} Relationship between $\Delta T_{\text{CDW}}$ and the number of \TTaS\ layers, where $\Delta T_{\text{CDW}}$ is defined as the difference in the CDW ordering temperatures between R1 and R2 in the 330 K -- 360 K range. The CDW ordering temperatures were calculated from the resistance inflections upon cooling and warming for devices D1--D4. \textbf{(d)} Schematic illustrating that few-layer \TTaS/\WSe\ heterostructures display a lower NC-CDW--IC-CDW transition temperature compared to \TTaS.  
    }
    \label{fig1}
\end{figure}

We fabricated mesoscopic devices from vertical heterostructures comprising \TTaS\ flakes of variable thickness and monolayer \WSe, which were encapsulated with hexagonal boron nitride (hBN) to prevent sample degradation and ensure pristine interfaces (Figure \ref{fig1}a). Each \TTaS\ crystal was partially interfaced with \WSe, allowing the creation of two distinct regions within each device: hBN/\ch{TaS2}/hBN (region R1) and hBN/\ch{WSe2}/\ch{TaS2}/hBN (region R2), as illustrated in Figure \ref{fig1}a. This device geometry allows us to directly probe the electronic influence of \WSe\ on \TTaS\ by comparing the transport properties of R1 and R2. We fabricated devices for electron transport consisting of 6-, 8-, 15- and 26-layer \TTaS, designated D1--D4, respectively, and subsequently measured their temperature-dependent longitudinal resistance ($R_{xx}$). Each device shows transport characteristics indicative of the transition from the metallic IC-CDW phase to the more insulating NC-CDW phase, including sharp resistance changes and a pronounced hysteresis in the cooling and warming profiles (Figure \ref{fig1}b,c). 

Our measurements reveal that CDW phase transitions above room temperature are substantially influenced by the presence of \WSe\ (Figure \ref{fig1}c). For the thinner devices, D1 and D2, the CDW ordering temperature is consistently higher for R1 (\TTaS\ region) compared to R2 (\TTaS/\WSe\ region), as seen in Figure \ref{fig1}c--d. This trend, observed across cooling and warming sweeps, suggests that interactions with \WSe\ induce CDW disorder in \TTaS\ and destabilize the more ordered NC-CDW phase \cite{yu2015gate,liu_superconductivity_2013}. On the other hand, for thicker devices (D3 and D4), a measurable difference between R1 and R2 is only observed upon warming, with R2 exhibiting a lower NC-CDW--IC-CDW ordering temperature (Figure \ref{fig1}c,d). This result is consistent with the introduction of CDW disorder owing to interactions between \WSe\ and \TTaS, with a pinning effect that is seemingly dependent on the CDW phase \cite{he2016distinct, ishiguro2020layer,Su2012collective, tsen2015structure,husremovic2023encoding}. We propose that structural defects in \TTaS\ brought on by the interface with \WSe\ may more strongly pin the NC-CDW state, which possesses frozen domains, compared to the ``melted'' IC-CDW phase, engendering a stronger CDW suppression upon warming \cite{husremovic2023encoding}. As a general observation, the CDW ordering temperature is increasingly affected with decreasing sample thickness, supporting a conclusion that local interface effects are driving the observed ensemble changes in \TTaS. Additionally, the presence of CDW modulation for D2, which possesses a large twist angle between \TTaS\ and \WSe, indicates that at least one factor contributing to suppression of CDW order is independent of the twist angle, as discussed further later.

\begin{figure}[!htbp]
    \centering
    \includegraphics[width=\textwidth]{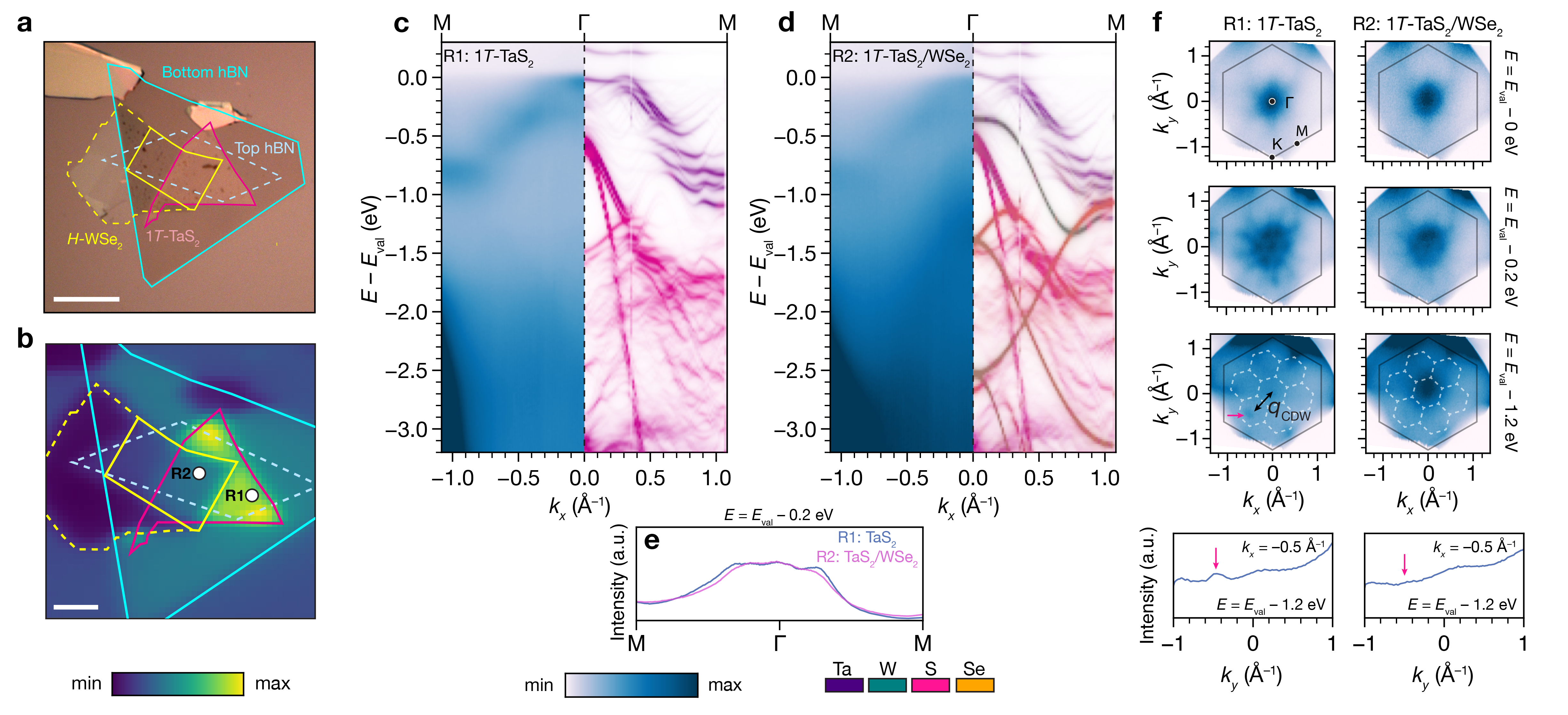}
    \caption[Electronic band structure of \TTaS\ and its heterostructure with \WSe.]{Electronic band structure of \TTaS\ and its heterostructure with \WSe. \textbf{(a)} Optical micrograph with the heterostrucutre components outline and marked. Note, the solid yellow line denotes monolayer WSe$_2$, while the dashed yellow line encloses the multilayer flake region. The sample was prepared on Si, and the twist angle between \TTaS\ and \WSe\ was estimated to be $\approx 2^\circ$ from the edge alignment of flakes in optical images. Scale bar: 10 $\mu$m. \textbf{(b)} Core-level map of the ARPES sample. The measured regions R1 and R2 are marked. These regions consist of (from top to bottom): R1 (monolayer hBN/9-layer \TTaS/$\sim$ 10 nm hBN) and R2 (monolayer hBN/monolayer \ch{WSe2}/9-layer \TTaS/$\sim$ 10 nm hBN). Color in the map corresponds to the summed intensity over the range from $-26.0$ eV to $-22$ eV, which spans the Ta $4f_{5/2}$ and $4f_{7/2}$ core levels. Scale bar: 5 $\mu$m. \textbf{(c,d)} Experimental (left) and calculated (right) ARPES band dispersions along the $\Upgamma$--M direction (top) for regions R1 \textbf{(c)} and R2 \textbf{(d)}. \textbf{(e)} Momentum distribution curves (MDCs) of R1 and R2 overlaid and normalized to intensity at $\Upgamma$. MDCs are centered at $E = E_{\mathrm{valence}} - 0.2$ eV and integrated over a 0.2 eV range. \textbf{(f)} Top: Isoenergy cuts for R1 and R2, binned over 0.2 eV. Solid black hexagons mark the normal-state Brillouin zone (BZ), dashed white hexagons the reconstructed BZ, and $q_\textit{cdw}$ the CDW wave vector. Bottom: Vertical MDCs at $E_{\mathrm{valence}} - 1.2$ eV, binned over $k_y = -0.4$ to $-0.6$ Å$^{-1}$ for R1 and R2. Pink arrows indicate signal at $q_\textit{cdw}$. Data in (c--f) were acquired at 20 K with h$\nu = 97.8$ eV and linear horizontal (LH) polarization.
    }
    \label{fig2}
\end{figure}

While electron transport measurements demonstrate a clear CDW modulation in the high temperature range, questions remain regarding the low-temperature electronic characteristics of \TTaS\ and its heterostructures with \WSe. Across all devices, we observe a significant increase in resistance with decreasing temperature below 150 K, indicating increasing CDW commensuration (Figure \ref{fig1}b) \cite{tsen2015structure,yoshida2014controlling,yu2015gate}. However, our measurements are limited to temperatures above 50 K due to the high resistance exhibited by the samples below this threshold. Additionally, the absence of hysteresis features in the low-temperature range for these thin flakes impedes a comparison of the electronic properties of R1 and R2 in the low-temperature phase (Figure \ref{fig1}b). To address these open questions, we prepared samples for nano angle-resolved photoemission spectroscopy (nanoARPES) measurements (Figure \ref{fig2}a) and measured them at 20 K. The submicron beam diameter in nanoARPES facilitates an independent assessment of the electronic band structure for sample regions R1 (hBN/\TTaS/hBN section) and R2 (hBN/\WSe/\TTaS/hBN section), as demonstrated in Figure \ref{fig2}b.

Figure \ref{fig2}c--d displays the ARPES band dispersion along the $\Upgamma$--M direction for R1 (Figure \ref{fig2}c) and R2 (Figure \ref{fig2}d). For R1, the electronic bands evince clear features of the strong electronic reconstruction in the presence of the $\sqrt{13}a \times \sqrt{13} a$ CDW superlattice: a flat band centered at $\Upgamma$ and opening of pseudogaps for the occupied band near the valence band energy $E_{\mathrm{valence}}$\cite{wang_band_2020,rossnagel2011origin,ritschel2015orbital,sohrt2014fast}. This is consistent with our band structure calculations, and previous studies of bulk \TTaS\ crystals \cite{wang_band_2020,rossnagel2011origin,ritschel2015orbital,sohrt2014fast}. The band structure of R2 is qualitatively similar to that of R1 but with increased blurriness (Figure \ref{fig2}d). Since nanoARPES is highly surface sensitive, we posit that this blurriness is due to the attenuation of the signal from \TTaS\ due to the overlaying \WSe. For a more quantitative comparison of R1 and R2, we examine their respective momentum distribution curves (MDCs) centered at $E= E_{\mathrm{valence}} - 0.2$ eV, summing over a 0.2 eV energy range (Figure \ref{fig2}e). The MDCs of R1 and R2 show similar features, but the heterostructure region exhibits increased band broadening between $\Upgamma$ and M. While some of these differences may result from the aforementioned signal attenuation, they are also consistent with a less pronounced CDW reconstruction in the heterostructure region.

\begin{figure}[!htbp]
    \centering
    \includegraphics[width=\textwidth]{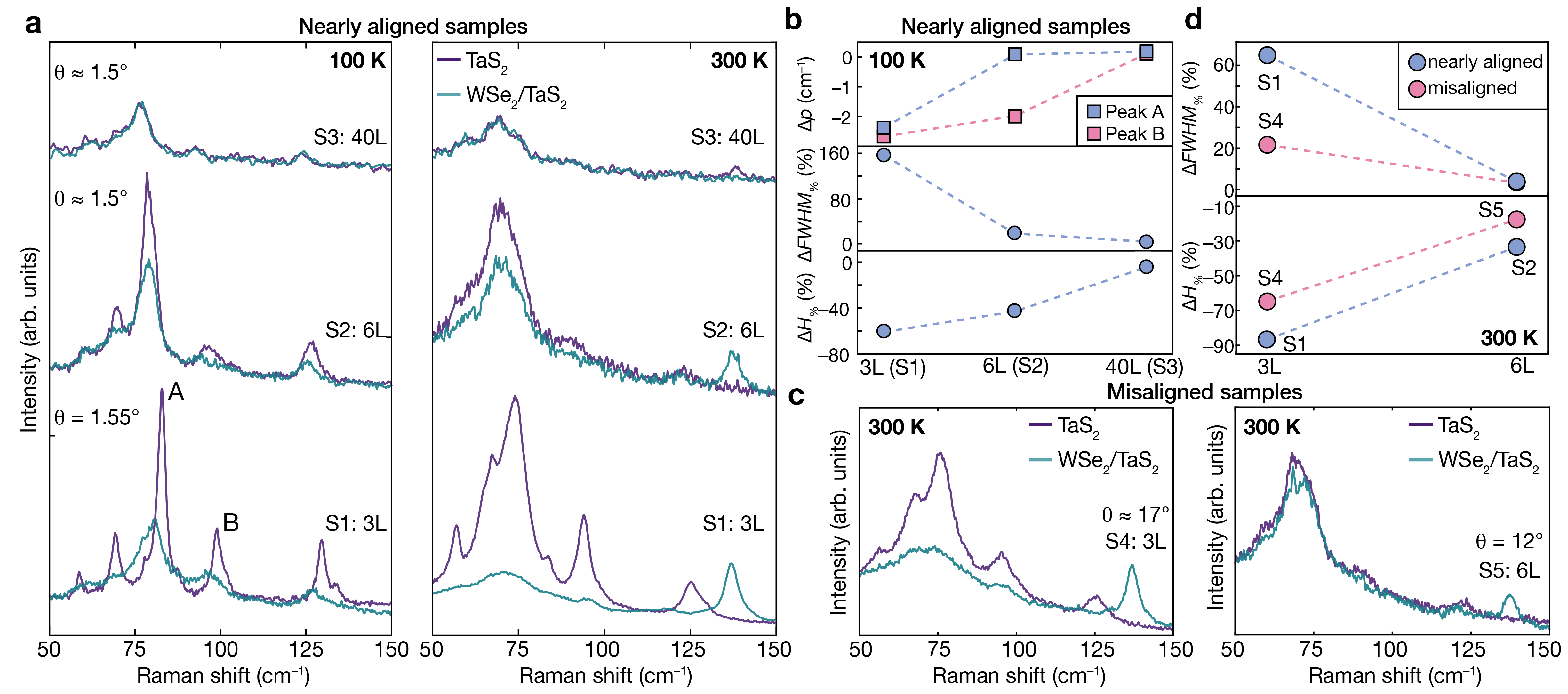}
    \caption[Thickness and twist dependence of CDW ordering in \TTaS/\WSe\ heterostructures] {Thickness and twist dependence of CDW ordering in \TTaS/\WSe\ heterostructures.
    \textbf{(a)} Raman spectra measured at 100 K (left) and 300 K (right) for nearly aligned \TTaS/\WSe\ heterostructures consisting of 3-, 6-, and 40-layer \TTaS\ (S1--S3). \textbf{(b)} Top: Raman peak shifts ($\Delta p$) of peaks A and B (labeled in a), calculated as the heterostructure peak position minus the \TTaS\ peak position for S1--S3 at 100 K. Middle: $\Delta$\textit{FWHM}${\%}$ and Bottom: $\Delta H{\%}$, based on analysis of peaks below 100 cm$^{-1}$ for S1--S3 at 100 K. \textbf{(c)} Raman spectra obtained at 300 K for azimuthally misaligned \TTaS/\WSe\ heterostructures (S4 and S5), with three and six layers of \TTaS, respectively. \textbf{(d)}  $\Delta$\textit{FWHM}$_{\%}$ and $\Delta H_{\%}$ for nearly aligned (S1, S2) and misaligned (S4, S5) samples at 300 K, based on the analysis of peaks below 80 cm$^{-1}$. Data points in (b,d) are derived from pseudo-Voigt fits of the measured Raman spectra.  For all samples, \TTaS\ flakes are partially interfaced with monolayer \WSe\ and encapsulated with hBN. Presented spectra are not normalized, and interlayer twist ($\theta$) is indicated for each sample. For S1 and S5, $\theta$ is determined from TEM diffraction, while for S2--S4, $\theta$ is determined from the alignment of flake edges in optical images. 
    }
    \label{fig3}
\end{figure}

To further investigate how heterostructuring affects the CDW state of \TTaS, we compare isoenergy cuts from regions R1 (standalone \TTaS) and R2 (heterostructure) at selected energies (Figure \ref{fig2}f). While the overall electronic structure is similar in both regions, notable differences appear at $E = E_{\mathrm{valence}} - 1.2$ eV. Specifically, region R1 exhibits additional electronic bands (highlighted by arrows in Figure \ref{fig3}f). These bands are located near the center of the reconstructed Brillouin zone and are rotated by approximately $13.9^{\circ}$ relative to the prominent sixfold-symmetric bands around $\Upgamma$ observed in both regions. This location and rotation are consistent with the expected signatures of CDW-induced band reconstruction \cite{yang2022visualization}. As follows, we propose that these features arise due to band folding from $\Upgamma$ into the CDW superstructure zone. Moreover, the increased prominence of these features in the standalone region (R1) hints that CDW commensuration may be stronger there compared to the heterostructure region (R2). Although this interpretation is tentative due to signal suppression in R2, we performed complementary optical measurements--Raman spectroscopy, photoluminescence, and reflection contrast--that probed the full sample volume to examine whether there are intrinsic differences in the CDW structure between the two regions.

\subsection{Examining CDW supression using Raman spectroscopy}
Next, we use Raman spectroscopy to probe the CDW suppression trends in \TTaS/\WSe\ heterostructures. Raman spectra reflect the CDW domain structure through distinct Raman-active modes arising from vibrations of the CDW lattice \cite{duffey1976raman,lacinska2022raman,yang2022visualization, albertini2016zone,he2016distinct}. Thus, changes in CDW ordering are typically evinced by modifications in the energy, broadness, and intensity of CDW-related Raman modes \cite{he2016distinct}. First, we investigate the Raman modes of three nearly aligned \TTaS/\WSe\ samples fabricated from 3-, 6-, and 40-layer \TTaS\ partially covered with monolayer \WSe\ (Figure \ref{fig3}a). These samples, labeled S1--S3, respectively, display noticeable differences in their temperature-dependent Raman spectra. Whereas the CDW modes of S3 appear unaffected by the presence of \WSe, the Raman spectra of S1 and S2 are substantially altered in contact with \WSe\ (Figure \ref{fig3}a). Specifically, for S1 and S2, CDW-related Raman modes in the heterostructure regions consistently exhibit broadening and attenuation compared to the \TTaS\ regions (Figure \ref{fig3}a). These characteristics were quantified by performing pseudo-Voigt fits to the data. Subsequently, to quantify peak height suppression, we sum the heights of CDW-related peaks for each sample (denoted as $H$) and compute the percent change upon stacking \WSe\ onto \TTaS\ ($\Delta H_{\%}$) as:

\begin{equation}
\Delta H_{\%} = \frac{H_{\text{TaS}_2} - H_{\text{heterostructure}}}{H_{\text{TaS}_2}} \times 100 \%
\end{equation}

We similarly calculate the percent change of the summed full width at half maximum (FWHM) of CDW-related peaks for each sample to assess changes in peak broadening, which reflects CDW disorder. This percent change ($\Delta$\textit{FWHM}$_{\%}$) is defined as:

\begin{equation}
\Delta \textit{FWHM}_{\%} = \frac{\textit{FWHM}_{\text{TaS}_2} - \textit{FWHM}_{\text{heterostructure}}}{\textit{FWHM}_{\text{TaS}_2}} \times 100 \%
\end{equation}

Evaluation of $\Delta H_{\%}$ and $\Delta$\textit{FWHM}$_{\%}$ for S1--S3 (Figure \ref{fig3}b, middle) reveals that peak broadening and height suppression diminish with increasing thickness. This is consistent with the attenuation of the Raman signal due to an interfacial interaction, which diminishes in significance for thicker samples. Importantly, all few-layer samples studied exhibit CDW suppression in the heterostructure region (Figure 3), and Raman mapping further confirms this suppression trend. Furthermore, a considerable redshift of up to 2.7 cm$^{-1}$ is observed for heterostructures at 100 K (Figure \ref{fig3}b, top), with the greatest redshifting of Raman modes measured for the thinnest (S1) sample. These observations align with our transport measurements: consistently stronger CDW suppression manifests in thinner samples. Having established that CDW suppression persists from 100 K to 300 K in nearly aligned heterostructures S1 and S2, we investigated azimuthally misaligned heterostructures S4 and S5, comprising 3- and 6-layer \TTaS, respectively (Figure \ref{fig3}c). In general, while the intensity attenuation and broadening of modes persisted in misaligned heterostructures, this effect was consistently more pronounced in nearly aligned heterostructures with the same number of layers (Figure \ref{fig3}d). This is consistent with a twist-dependent characteristic of CDW suppression, possibly stemming from the evolution of charge transfer, hybridization, or moir\'e strain due to varying interlayer alignment.

\subsection{Effects of charge transfer and moir\'e strain on the CDW suppression}
\begin{figure}[!htbp]
    \centering
    \includegraphics[width=\textwidth]{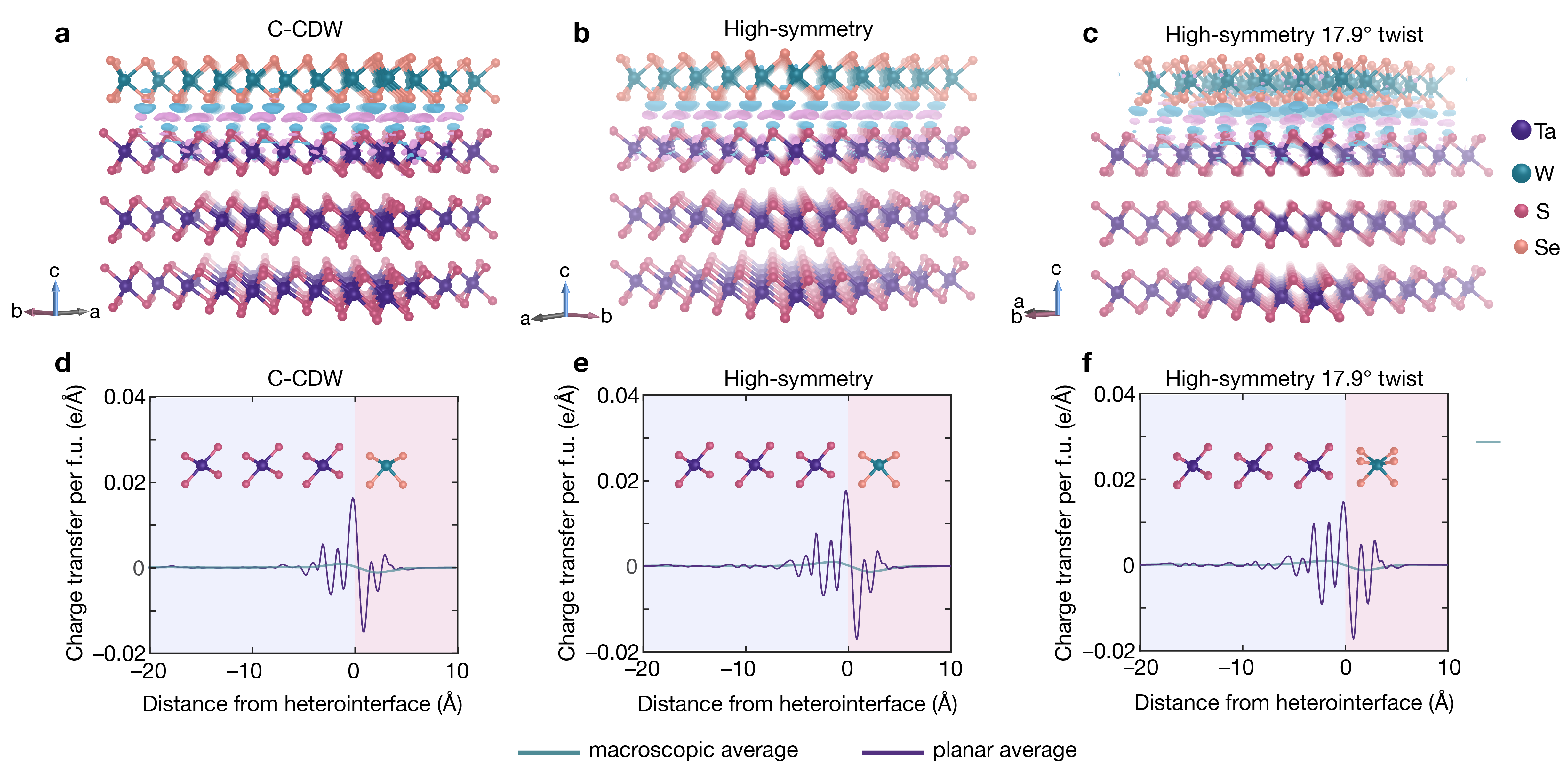}
    \caption[Local charge transfer characteristics across CDW phases and twist angles.] {Local charge transfer characteristics across CDW phases and twist angles. Charge transfer (top) and the corresponding planar and macroscopic averages per formula unit (bottom) for 3-layer \TTaS\ interfaced with monolayer \WSe\ for: an azimuthally nearly aligned heterostructure in the C-CDW phase \textbf{(a)}, azimuthally nearly aligned heterostructure in the high-symmetry metallic phase \textbf{(b)}, and a twisted interface with 17.9$^{\circ}$ misalignment between the high-symmetry \TTaS\ and \WSe\ layers  \textbf{(c)}. The pink and blue regions in the charge transfer plots (top) represent charge accumulation and depletion, respectively. Isosurface cutoff was set to 0.0003 eV for (a,b) and 0.0004 eV for (c). 
    }
    \label{fig4}
\end{figure}
Next, we examine the cause of the general trend of CDW suppression observed across both nearly aligned and misaligned heterostructures. Band structure calculations reveal minimal effects of hybridization, even in nearly aligned heterostructures where such effects should be most pronounced. We conclude that hybridization cannot account for the observed CDW suppression in \TTaS/\WSe\ heterostructures. Next, we investigate whether charge transfer between \TTaS\ and \WSe\ could explain the CDW suppression trends. Density functional theory (DFT) calculations were performed on heterostructures composed of three layers of \TTaS\ and a monolayer of \WSe. Three configurations were considered: (1) C-CDW in \TTaS\ layers (Figure \ref{fig4}a), (2) high-symmetry \TTaS\ with no charge density wave (Figure \ref{fig4}b), and (3) a twisted interface with 17.9$^{\circ}$ between high-symmetry \TTaS\ and \WSe\ layers (Figure \ref{fig4}c). Across all configurations, the calculated charge transfer characteristics are similar (Figure \ref{fig4}a--c). Specifically, only the singular \TTaS\ layer directly interfaced with \WSe\ is electron-doped, while \WSe\ is hole-doped. Given that charge transfer is evident in all calculated configurations and does not vary significantly with twist angle or CDW phase, we propose that this charge modulation at the \WSe\ interface is responsible for the overall suppression trend observed in both nearly aligned and misaligned heterostructures.

Interestingly, although charge transfer is confined to a singular \TTaS\ layer (Figure \ref{fig4}), this effect appears to impact the ensemble CDW ordering of the entire sample (Figures \ref{fig1}--\ref{fig3}). This local charge transfer alters the ionicity of the layers involved, decreasing electron density at the bonds and increasing it at the ionic sites (Figure \ref{fig4}a--c). These changes in ionicity may modulate electron--phonon coupling\cite{meijerink1996electron}, disrupting CDW ordering within the single layer, which then impacts global ordering due to the strong interlayer interactions in \TTaS \cite{robbins1980x, ritschel2015orbital, stahl2020collapse,scruby1975role,tanda1984x,butler2020mottness,ritschel2018stacking}. Taken together, we propose that local charge transfer contributes to the overarching trend of globally increased CDW suppression in \TTaS/\WSe\ heterostructures. However, these charge transfer calculations do not explain why CDW suppression is more pronounced in nearly aligned samples compared to misaligned ones (Figure \ref{fig3}d).

\begin{figure}[!htbp]
    \centering
    \includegraphics[width=5.4in]{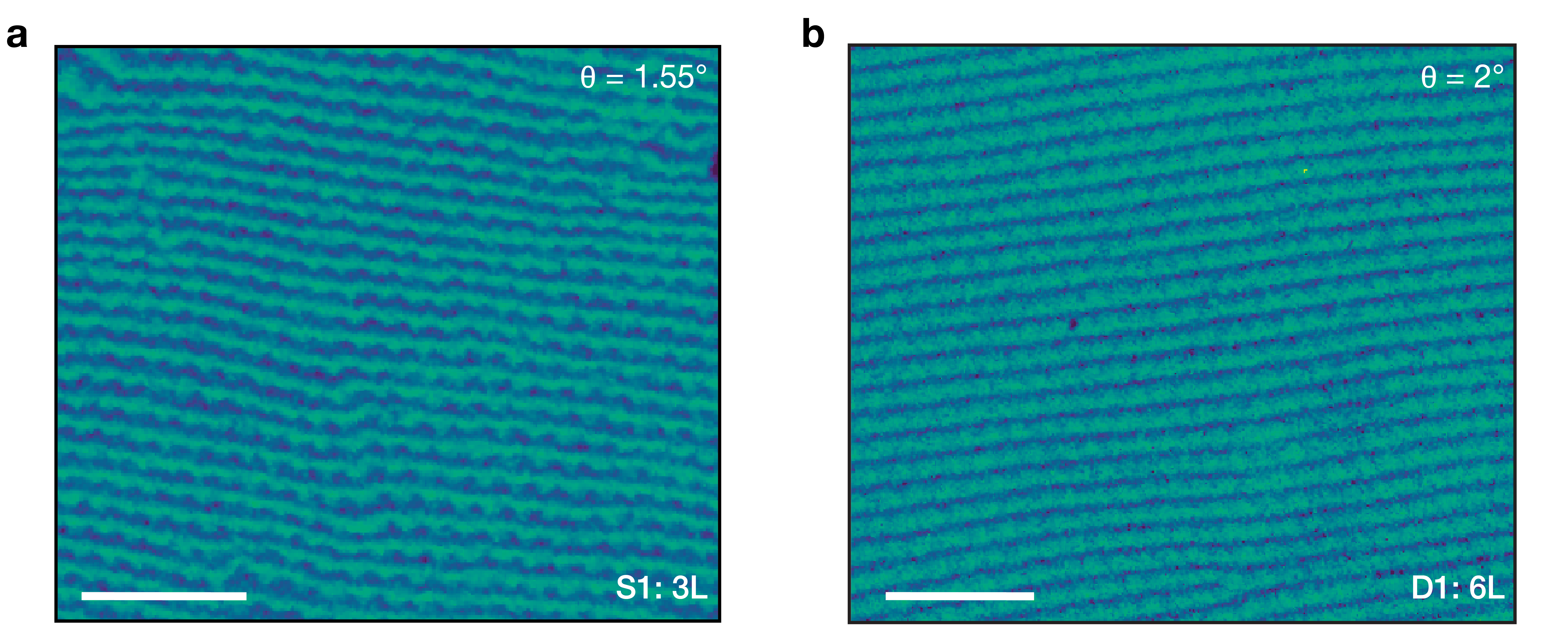}
    \caption[Lattice reconstruction of nearly aligned samples.] {Lattice reconstruction of nearly aligned samples. \textbf{(a,b)} Dark-field transmission electron microscopy (TEM) lattice reconstruction of nearly aligned samples S1 \textbf{(a)} and D1 \textbf{(b)}. Micrographs were constructed from the closest first-order peaks of \TTaS\ and \WSe. Scale bars: 50 nm. 
    }
    \label{fig5}
\end{figure}

To understand the effect of alignment, we investigate whether moir\'e relaxation occurs between \TTaS\ and \WSe\, as the resultant strain could account for the increased CDW suppression in nearly aligned samples. We examine the lattice reconstruction of D1 and S1 using dark-field TEM (Figure \ref{fig5}). The dark-field micrographs reveal intensity variations that are indicative of reconstructed lattice domains (Figure \ref{fig5})\cite{yoo_atomic_2019,weston_atomic_2020,kazmierczak_strain_2021, VanWinkle2023}. In contrast, moir\'e reconstruction is not expected in misaligned samples with large twist angles. These TEM data support the notion that moir\'e reconstruction and the accompanying strain may explain the twist-dependent CDW properties in heterostructures of \TTaS\ and \WSe. We note that the small lattice mismatch of 2$\%$ between \TTaS\ and \WSe\ enables the observation of lattice reconstruction even at negligible interlayer twist angles, primarily driven by lattice dilations in heterobilayers\cite{VanWinkle2023}. Lattice dilations, defined as volumetric strain, are local changes in volume compared to a rigid moir\'e. They manifest as periodic stretching and expansion of layers, creating high-energy and low-energy regions, corresponding to higher and lower effective lattice mismatches, respectively \cite{VanWinkle2023}. Consequently, since lattice reconstructions can introduce periodic strain fields, they might be expected to influence the ordering of CDWs and lead to increased CDW suppression in small-twist-angle samples \cite{zhao_moire_2021}. Thus, combining theory and experimental data, we propose that charge transfer contributes to an overarching trend of increased CDW disorder in \TTaS/\WSe\ heterostructures, while moir\'e relaxation strain seemingly amplifies this suppression in nearly aligned samples.

\subsection{Effects of heterostructuring on the optical properties of \WSe}
\begin{figure}[!htbp]
    \centering
    \includegraphics[width=\textwidth]{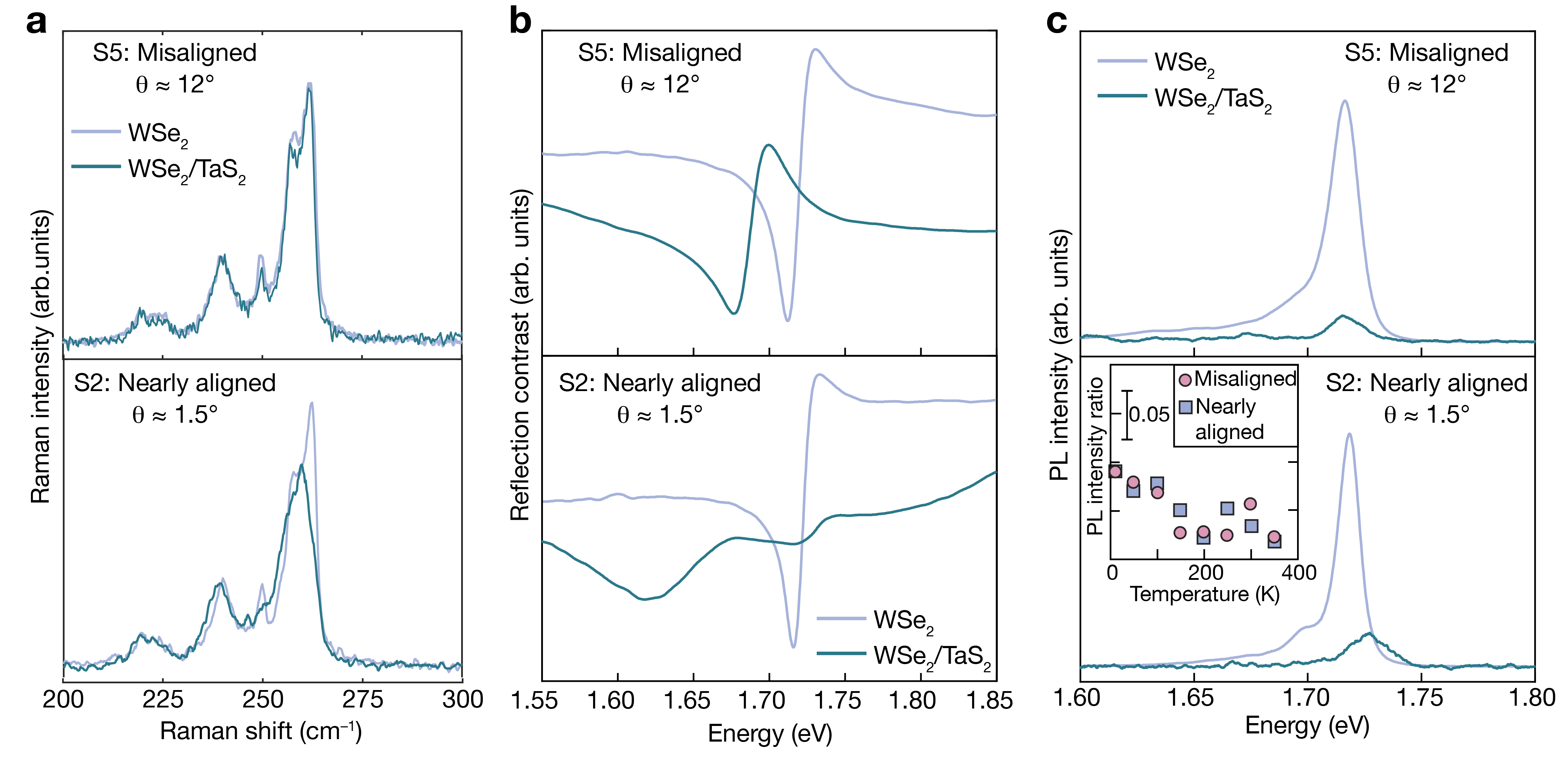}
    \caption[Effects of heterostructuring on the optical properties of \WSe]{Effects of heterostructuring on the optical properties of \WSe. \textbf{(a--c)} Raman spectra \textbf{(a)}, reflection contrast \textbf{(b)}, and photoluminescence \textbf{(c)} for the azimuthally misaligned heterostructure S5 (top) and the azimuthally nearly aligned heterostructure S2 (bottom). The inset in (c) displays the temperature-dependent PL intensity ratio of the nearly aligned (S2) and misaligned (S5) samples. This ratio is obtained by dividing the integrated intensity of the $A_{\mathrm{1s}}$ exciton in the heterostructure region by that in the \WSe\ region. Raman spectra were acquired with the 633 nm laser at 300 K, while PL and RC were obtained with a 532 nm laser at 100 K.}
    \label{fig6}
\end{figure}

Next, we investigate the effects of heterostructuring on the Raman modes of \WSe. Figure \ref{fig6}a displays Raman spectra for nearly aligned (S2) and misaligned (S5) samples, while corresponding CDW-related Raman modes were previously shown in Figure \ref{fig3}a,c. In the azimuthally misaligned sample (S5), Raman modes of \WSe\ exhibit minimal changes between the standalone and heterostructure regions, as confirmed by Lorentzian peak analysis. In contrast, Raman modes of \WSe\ are strongly modified in the heterostructure region of the nearly aligned sample (S2). Lorentzian fits to the data reveal that the 250 cm$^{-1}$ Raman mode, comprising degenerate $E'/A_{1}'$ modes in standalone \WSe \cite{terrones2014new,shi2016raman,delcorro_excited_2014,dadgar2018strain}, splits into two distinct peaks at 246 cm$^{-1}$ and 250 cm$^{-1}$ in the heterostructure region. Additionally, higher-energy peaks above 250 cm$^{-1}$, encompassing $2LA(M)$, $A_{2}''$, and $A_{1}'(M)$ modes\cite{terrones2014new,shi2016raman}, exhibit red-shifts and broadening.

Generally, the in-plane $E$ modes in transition metal dichalcogenides (TMDs) are sensitive to moir\'e strain\cite{lv2023strain,rahman2022extraordinary,quan2021phonon,roy2023uniaxial,dadgar2018strain}, whereas the out-of-plane $A$ modes respond weakly to strain but strongly to doping\cite{quan2021phonon,liu2014evolution,dadgar2018strain,debnath2020evolution,fan2020tailoring}. While our calculations indicate charge transfer between \TTaS\ and \WSe, the magnitude of this charge transfer is small ($\sim$ 0.01 electrons per formula unit of \WSe), so we do not expect large shifts of the $A_{1}'$ mode. Thus, we assign the peak remaining at 250 cm$^{-1}$ to the $A_{1}'$ mode and the red-shifted 246 cm$^{-1}$ peak to the $E'$ mode. These spectral features are consistent with the strain effects previously reported in monolayer \WSe\cite{dadgar2018strain}. The red shifting and broadening of higher-energy modes is also congruent with effects of strain \cite{dadgar2018strain}.

Importantly, Raman trends observed in 6L samples match those in 3L samples, indicating systematic rather than extrinsic or unintentional strain effects. Therefore, the observed Raman trends are congruent with the modification of the \WSe\ phonon modes by moir\'e relaxation and strain.

We next examine how the optoelectronic characteristics of \WSe\ are affected by the presence of \TTaS. To this end, reflection contrast (RC) was measured for S2 and S5. For both samples, significant broadening of the RC signal was detected in the heterostructure region compared to \WSe\ (Figure 6b). According to the uncertainty principle, this spectral broadening reflects the timescale of the charge-transfer process and is consistent with a shortened exciton lifetime due to ultrafast charge transfer to \TTaS\ \cite{hill_broadening_2017, rigosi_broadening_2015} (Supplementary Note 1). Although broadening is observed in both nearly aligned (S2) and misaligned (S5) samples, quantitative analysis of the RC signal shows that the $A_{\mathrm{1s}}$ exciton of \WSe\ is significantly broader in the nearly aligned heterostructure. Specifically, we measure a broadening of $\sim$4 meV for the misaligned sample S5, compared to $\sim$10 meV for the nearly aligned sample S2. This increased broadening indicates a faster charge-transfer rate in the nearly aligned case. Additionally, the larger attenuation of the RC intensity observed for the nearly aligned sample corresponds to a greater reduction in exciton oscillator strength, which further supports faster and more efficient charge transfer. This likely arises from improved interlayer orbital overlap at smaller twist angles. Interestingly, RC of the nearly aligned sample (S2) contains an additional peak below the $A_{\mathrm{1s}}$ exciton energy at all temperatures (Figure \ref{fig6}b), indicating strong light absorption around these energies. Further investigation is needed to clarify the origin of this peak in the aligned sample.

To further delineate the effect of \TTaS\ on \WSe, we examine the photoluminescence spectra (PL) of S2 and S5. The PL intensity of $A_{\mathrm{1s}}$ exciton of \WSe\ is significantly quenched in both nearly aligned and misaligned heterostructures (Figure \ref{fig6}c). This can be explained by charge transfer from \WSe\ to \TTaS\ where the photoexcited electrons and holes in \WSe\ relax into lower energy states provided by \TTaS. Such charge transfer in vdW heterostructures has been shown to occur on an ultrafast timescale (sub-ps)\cite{shrestha_interfacial_2023} relative to the radiative lifetime of excitons, facilitating efficient quenching of the PL. In the case of \TTaS/\WSe, band structure calculations indicate that energy minima in \TTaS\ ($\Upgamma$) and \WSe\ (K) are mismatched in momentum. Hence, such an indirect nature of charge transfer across \TTaS/\WSe\ correlates with the insensitivity of PL quenching to the twist angle (Figure \ref{fig6}c, inset). A small deviation from the PL intensity ratio trend is observed for the nearly aligned and misaligned heterostructures at 250 K and 300 K, respectively. This deviation may stem from changes in Coulomb screening associated with CDW transitions, but further investigation is needed to probe and elucidate this trend.

The nearly aligned and misaligned heterostructures also exhibit differences in $A_{\mathrm{1s}}$ exciton energy relative to the isolated \WSe\ as seen from both PL and RC spectra (Figure \ref{fig6}b,c). Such a variation in the exciton peak position in vdW heterostructures may arise from the net change of electronic band gap and exciton binding energy due to dielectric screening \cite{raja_coulomb_2017}, charge doping \cite{chernikov_doping_2015}, interlayer coupling \cite{liu2014evolution}, and strain \cite{lv2023strain}. The $A_{\mathrm{1s}}$ exciton energy shows a dominant red shift in the RC spectra on the misaligned sample. Strong screening of Coulomb interactions in \WSe\ from the charge density within \TTaS\ causes a reduction in the quasiparticle bandgap and exciton binding energy, resulting in a net red-shift in the $A_{\mathrm{1s}}$ exciton energy. Despite an identical dielectric screening effect expected in the nearly aligned heterostructure, an evident blue-shift of exciton energy in both RC and PL indicates an interplay of other mechanisms.  We do not expect a strong modulation in the electronic coupling (i.e. band hybridization) with twist angle since our band structures reveal that there is no hybridization of \TTaS\ with \WSe\ at the K point, where the conduction band maximum and valence band minimum of \WSe\ are located. Inferences of similar charge transfer and doping for different twist angles from the calculations and PL quenching hint at the role of moir\'e strain \cite{lv2023strain,rahman2022extraordinary} and the corresponding band gap renormalization \cite{liu2014evolution} from the moir\'e superlattice in the nearly aligned sample.

\section{Conclusion}
Here, we investigate the optoelectronic implications of heterostructuring \TTaS\ with monolayer \WSe. We examine heterostructures with varying numbers of \TTaS\ layers and interlayer alignments, revealing that the CDW is suppressed upon contact with \WSe, with this effect being enhanced by decreasing sample thickness and minimizing the interlayer twist angle. The IC-CDW to NC-CDW transition is strongly influenced by the presence of \WSe\ in thin samples, as evidenced by electron transport measurements. Moreover, nanoARPES confirms strong band renormalization in ultrathin \TTaS\ samples and the influence of \WSe\ presence on the bands of \TTaS. Raman spectroscopy systematically reveals evidence of CDW disorder in \TTaS/\WSe\ heterostructures, which is enhanced in nearly aligned samples. These nearly aligned samples were analyzed with dark-field TEM, revealing evidence of twist-dependent moir\'e reconstruction. Correlating these experimental observations with theoretical insights, we deduce that strain resulting from lattice reconstructions is the most plausible explanation for the increased CDW suppression observed in nearly aligned heterostructures. We also find that charge transfer, irrespective of twist, consistently leads to CDW suppression in thin samples contacting \WSe. Density functional theory (DFT) elucidates that this charge transfer is localized to the \TTaS\ layer directly interacting with \WSe, underscoring the significance of out-of-plane interactions in CDW ordering of \TTaS. 

We also observe modifications in the optoelectronic properties of \WSe\ upon contact with \TTaS. Specifically, the broadening of RC and quenching of PL point to rapid exciton dissociation dynamics in contact with \TTaS. Moreover, we measure twist-dependent shifts in PL and RC, coupled with increased broadening and emergence of a new peak in RC of nearly aligned samples. These data suggest that moir\'e effects may alter exciton energies and dissociation dynamics. This study demonstrates that interfacing CDW crystals and semiconductors may be a viable framework to simultaneously tune their respective electronic properties, enabling the development of optoelectronic device schemes that couple optical manipulation with CDW phase transitions.

\section{Methods}

\subsection{hBN exfoliation}
Chips of SiO$_2$/Si were cut into $\sim$ $1 \times 1$ cm squares and subjected to ozone cleaning for $\sim$ 60 minutes at 150 $^{\circ}$C. Immediately before concluding the cleaning, 2--4 hexagonal boron nitride (hBN) crystals (received from T. Taniguchi and K. Watanabe), were tessellated using Scotch tape. The SiO$_2$/Si chips were then removed from the ozone cleaner and pressed onto the tape for 10 minutes with finger pressure before removing the tape from the chips. 

\subsection{\TTaS\ exfoliation}
First, 90 nm SiO$_2$/Si chips were plasma cleaned for 2--3 minutes. Next, the chips were heated on a hotplate in an Ar glovebox for $\sim$ 1 hour. Before completing the cleaning, a $\sim 2 \times 2$ mm piece of a \TTaS\ crystal (HQ Graphene) was tessellated using Scotch tape. The chips were then pressed onto the tape for 10 minutes with finger pressure and then quickly peeled off the tape to yield thin flakes on the substrate. Flake thickness was determined by optical contrast versus the SiO$_2$/Si substrate \cite{li2013rapid}.

\subsection{2\textit{H}-\ch{WSe2} exfoliation}
First, 285 nm SiO$_2$/Si chips were cleaned using ozone for $\sim$ 30-60 minutes. Next, the chips were heated on a hotplate in an Ar glovebox at 200 $^{\circ}$C for at least 1 hour. Before concluding the cleaning, a $\sim 2\times 2$ mm piece of 2\textit{H}-\ch{WSe2} (HQ graphene) was tessellated using Magic Scotch tape and pressed onto a piece of Gel-Pak 4 polydimethysiloxane (PDMS) film secured on a glass slide. Following the heating, the chips were pressed onto the PDMS for $\sim$ 30 seconds and then lifted off from the PDMS. 

\subsection{Fabrication of bottom contacts}
Bottom contacts were patterned using electron-beam lithography (Crestec CABL-UH Series Electron Beam Lithography System) onto atomically flat, 10--30 nm-thick hBN flakes. Initially, the contacts did not extend past the edge of the hBN crystals. Next, 7--10 nm of metal (Pt or Au) was evaporated onto the lithographically defined contacts, which were subsequently lifted off in acetone overnight. Following stacking, electron-beam lithography and thermal evaporation were used to pattern 0.5 nm Cr/100 nm Au extended contacts. 

\subsection{Fabrication of vdW heterostructures}
Stacking was performed in an Ar-filled glovebox using the dry transfer method with a PC/PDMS stamp \cite{husremovic2022hard}. The \TTaS\ flakes were always exfoliated no more than 2 days before stacking to prevent sample degradation. All samples for Raman studies were prepared on 90 nm SiO$_2$/Si substrate, which enhances the weak optical contrast signal from \TTaS.

To determine the interlayer twist angle between \WSe\ and \TTaS, we aligned their zigzag edges. Here, 0$^\circ$ twist corresponds to the configuration in which the zigzag edges of \WSe\ and \TTaS\ are parallel.

Zigzag edges were identified optically based on flake edge angles. In prior work\cite{husremovic2022hard,husremovic2023encoding,husremovic_tailored_2025}, we performed cross-sectional TEM on over a dozen TaS$_2$ flakes (including both 1\textit{T} and 2\textit{H} polymorphs) and found that edges forming 60$^\circ$ or 120$^\circ$ angles consistently correspond to the zigzag direction. The same trend has been reported for MoS$_2$\cite{guo_distinctive_2016}, where 60$^\circ$ edges reliably indicate the zigzag orientation, with accuracy comparable to SHG measurements. For samples where interlayer twist could be independently measured, we found that the twist angles estimated optically using this method typically agreed within 2$^\circ$ of values obtained from SHG or TEM diffraction, supporting the reliability of edge-based alignment.

\subsection{Sample preparation for transmission electron microscopy (TEM).}  
Samples were prepared in an Ar-filled glovebox using the dry transfer method\cite{husremovic2022hard,husremovic2023encoding}. A transfer stamp composed of a poly(bisphenol A carbonate) (PC) film on polydimethylsiloxane (PDMS), supported by a glass slide, was used to pick up flakes at 140 $^{\circ}$C--150 $^{\circ}$C. The heterostructures were then transferred onto 200 nm silicon nitride holey TEM grids (Norcada), which were treated with O$_2$ plasma for 5 minutes immediately prior to stacking. The PC was melted onto the grid at 150\,$^\circ$C and dissolved in chloroform for 1 hour.

\subsection{Electron transport measurements}
A constant alternating current of 0.1--0.5 $\mu$A (17.777 Hz) was applied between source and drain contacts and 4-probe $V_{xx}$ was measured using a lock-in amplifier (Stanford Research SR830). Measurements were collected between pairs of contacts ensuring individual assessment of R1 and R2. All measurements were performed in a Quantum Design Physical Property Measurement System (PPMS). Due to the relatively large bandgap of \WSe\ compared to \TTaS, \WSe\ does not contribute significantly to the measured transport behavior. 

\subsection{Raman spectroscopy}
Raman spectroscopy (Horiba Multiline LabRam Evolution) was conducted using 633 nm continuous wave laser with an ultra-low frequency filter. Spectra were typically collected with 50–80 $\mu$W power and an 1800 gr/mm grating, with 20-second acquisition times and 2--4 accumulations. For temperature-dependent measurements, samples were measured in an optical cryostat (Cryo Industries of America), which was evacuated to $\sim 10^{-6}$ Torr and cooled with liquid nitrogen. The samples were placed with thermally conductive vacuum grease onto a copper puck, which is temperature-controlled using resistive heating (LakeShore Cryotronics Model 325). Due to fluctuations in the system, temperatures reported throughout are accurate to $\pm$ 5 K. The laser beam was focused on the sample into a spot size of approximately 1 $\mu$m. 

\subsection{Photoluminescence and reflection contrast}
Temperature-dependent Photoluminescence (PL) and Reflection contrast (RC) spectroscopy were performed on the samples mounted in a Montana Cryostation S200. For the PL measurements, a 532 nm diode-pumped solid-state laser (Coherent$^{TM}$ Sapphire SF NX) is used as the excitation source. For the RC measurements, a Quartz Tungsten Halogen Research Light Source (Oriel instruments) is used as the broadband excitation source. The excitation beam is focused ($<1\mu$m dia) onto the sample loaded on the Agile Temperature Sample mount (ATSM) with a 100x objective (NA = 0.9). The emission or reflection signal from the sample is collected in back-scattering geometry and dispersed with AndorTM Kymera 328i spectrometer using 300 l/mm grating onto the detector (TE-cooled Andor$^{TM}$ Newton DU970P EMCCD). The laser excitation power on the sample was kept below 1 $\mu$W.

\subsection{Transmission electron microscopy} 
Selected area electron diffraction (SAED) patterns, used for determining interlayer twist, were obtained with a 40 $\mu$m diameter aperture (defining a selected diameter of $\sim$720 nm on the sample) using FEI TitanX TEM operated at 60--80 kV. Dark-field TEM was also obtained using FEI TitanX TEM with a 10--20 $\mu$m objective aperture. 

\subsection{Density functional theory}
Density functional theory (DFT) calculations were carried out on heterostructures composed of three layers of 1\textit{T}-\ce{TaS2} and one layer of 1\textit{H}-\ce{WSe2}. Three cases were investigated: a commensurate charge density wave in the \ce{TaS2} layers (CCDW); high-symmetry 1\textit{T}-\ce{TaS2} with no charge density wave (HS); and a twisted interface with 17.9$^{\circ}$ between high-symmetry \ce{TaS2} and \ce{WSe2} layers. The twisted heterostructure was generated using the \textsc{twister} code \cite{naik_twister_2022,naik_ultraflatbands_2018}. Structures had AA stacking with all metal centers nearly aligned except across the twisted interface. In all cases, the vacuum region between slabs was greater than \SI{35}{\angstrom}.

\subsection{Nano angle-resolved photoemission spectroscopy (nanoARPES)}
nanoARPES data were collected at Beamline 7.0.2 of the Advanced Light Source (ALS) on the nanoARPES endstation using Scienta Omicron R4000 hemispherical electron analyzers. The beam diameter was approximately $1$ $\mu$m. All measurements were conducted at the base temperature of 20 K with linear horizontal (LH) polarization at pressures lower than $5 \times 10^{-11}$ Torr. Data analysis was conducted using the PyARPES software package \cite{stansbury2020pyarpes}. 

\subsection{Spin-polarized density functional theory calculations}
Spin-polarized DFT calculations were carried out using the Vienna \textit{Ab initio} Simulation Package (\textsc{vasp}) \cite{Kresse1993,Kresse1994,Kresse1996,Kresse1996a} with projector augmented wave (PAW) pseudopotentials \cite{blochl1994,Kresse1999} including Ta 5\textit{pd}6\textit{s}, W 5\textit{spd}6\textit{s}, S 3\textit{sp} and Se 4\textit{sp} electrons as valence. The plane-wave energy cutoff was set to 400 eV and the \textit{k}-point grids were Gamma-centred with a \textit{k}-point spacing less than \SI{0.2}{\per\angstrom}, which gave an energy convergence of 1 meV per atom. The convergence criteria for the electronic self-consistent loop was set to 10\textsuperscript{-6} eV.

A Hubbard-$U$ correction was applied to capture the charge distribution in the CCDW. A $U$ of \SI{2.86}{\electronvolt} was applied within the Dudarev approach~\cite{Dudarev1998} to the $d$ orbitals of the Ta located at the center of the Star of David distortion, following the self-consistent calculation of $U$ in \cite{boix-constant_out_2021}. This captures a band gap opening in \ce{TaS2} between the spin-polarized bands at the Fermi level, in contrast to the metallic HS structure.

Structural optimizations were done using the PBEsol \cite{Perdew2008} exchange-correlation functional until the residual forces on the ions were less than \SI{0.001}{\electronvolt\per\angstrom} for the CCDW and HS structures, whereas the twisted structure was fixed with van der Waals gaps set to those of the optimized HS structure. The percentage strain in the constituent layers of the heterostructures is given by comparison to the lattice parameters of individually optimized \ce{TaS2} trilayers and \ce{WSe2} monolayer in Table \ref{table}.

\begin{table}[!htbp]
\centering
\caption[DFT parameters for \TTaS/\WSe\ heterostructures.]{DFT parameters for \TTaS/\WSe\ heterostructures. In-plane lattice parameter ($a$) and percentage strain for the heterostructures compared to the individual constituents. All structures are optimized for a monolayer of \ce{WSe2} and a trilayer of \ce{TaS2}.}
\label{table}
\begin{tabular}{@{}lcccc@{}}
\toprule
Structure                    & $a$ (Å) & \multicolumn{2}{c}{$\Delta a/a$ (\%)}                                            &  \\
                             &                         & \ce{TaS2}  & \ce{WSe2} &  \\  \midrule
\ce{WSe2}                      & 3.273                   &                                   &                                   &  \\
HS-\ce{TaS2}                   & 3.326                   &                                   &                                   &  \\
HS-\ce{TaS2} | \ce{WSe2}         & 3.312                   & -0.41                             & 1.21                              &  \\
CCDW-\ce{TaS2}                 & 12.012                  &                                   &                                   &  \\
CCDW-\ce{TaS2} | \ce{WSe2}       & 11.955                  & -0.48                             & 1.30                              &  \\
Twisted HS-\ce{TaS2} | \ce{WSe2} & 18.441                  & -0.42                             & 1.20                              & \\  \bottomrule
\end{tabular}
\end{table}

Orbital-projected band structures were plotted using the \textsc{sumo} package \cite{ganose_sumo_2018}, and unfolded band structures using \textsc{easyunfold} \cite{zhu_easyunfold_2024}. Coloring is applied to bands with a magnitude relative to orbital decomposition on atomic sites based on projector functions within the PAW method in VASP.

Charge transfer plots were generated by subtracting the charge densities of individual \ce{TaS2} and \ce{WSe2} fragments from the charge density of the heterostructure. No structure relaxation was performed for the individual fragments; atom positions were identical to those in the optimized heterostructure. This charge density difference was summed within planes perpendicular to the $c$-axis parallel and normalized per formula unit within a layer. The macroscopic average of this 1D charge difference was calculated as a running average performed three times over periods of \SI{3.03}{\angstrom}, to smooth over the periodic features of the structure. 1D plots were calculated using the \textsc{vaspkit} code \cite{wang_vaspkit_2021}. Charge density difference was visualized using \textsc{vesta} \cite{momma_vesta_2008}.


\section*{Acknowledgements}
This material is based upon work supported by the US National Science Foundation Early Career Development Program (CAREER), under award no. 2238196 (D.K.B). Work at the Molecular Foundry, LBNL, was supported by the Office of Science, Office of Basic Energy Sciences, the U.S. Department of Energy under Contract no. DE-AC02-05CH11231.  This research used resources of the Advanced Light Source, which is a DOE Office of Science User Facility under contract no. DE-AC02-05CH11231. Confocal Raman spectroscopy was supported by a Defense University Research Instrumentation Program grant through the Office of Naval Research under award no. N00014-20-1-2599 (D.K.B.). L.S.X. was supported by an Arnold O. Beckman Postdoctoral Fellowship. S.H. acknowledges support from the Blavatnik Innovation Fellowship. K.W. and T.T. acknowledge support from the JSPS KAKENHI (Grant Numbers 21H05233 and 23H02052) , the CREST (JPMJCR24A5), JST and World Premier International Research Center Initiative (WPI), MEXT, Japan. K.I. was supported by the UK Engineering and Physical Sciences Research Council (EPSRC) through grant No. EP/W028131/1. DFT computations were performed using the Sulis Tier 2 HPC platform funded by EP/T022108/1 and the HPC Midlands+ consortium. Through membership of the UK’s HEC Materials Chemistry Consortium, EP/X035859, this work also used ARCHER2 UK National Supercomputing Services.

\section*{Competing Interests}
The authors declare no competing interests.

\section*{Additional Information}
Correspondence and requests for materials should be emailed to DKB\\
(email: bediako@berkeley.edu).

\printbibliography
\end{document}